\documentclass[fleqn,usenatbib,referee]{mnras}

\usepackage[T1]{fontenc}

\DeclareRobustCommand{\VAN}[3]{#2}
\let\VANthebibliography\thebibliography
\def\thebibliography{\DeclareRobustCommand{\VAN}[3]{##3}\VANthebibliography}

\usepackage{graphicx}

\usepackage{amsmath}    
\usepackage{amssymb}    

\title[On the need of an ultramassive black hole in OJ 287]{On the need of an ultramassive black hole in OJ 287}

\author[M. Valtonen et al.]{Mauri J. Valtonen,$^{1,2}$\thanks{E-mail: mvaltonen2001@yahoo.com (MJV)} 
Staszek Zola,$^3$ A. Gopakumar,$^4$
Anne L\"ahteenm\"aki,$^{5,6}$
\newauthor{Merja Tornikoski,$^5$ Lankeswar Dey,$^4$ Alok C. Gupta,$^{7,20}$ Tapio Pursimo,$^8$ Emil Knudstrup,$^8$}
\newauthor{Jose L. Gomez,$^9$ Rene Hudec,$^{10,11}$ Martin Jel\'{\i}nek,$^{11}$ Jan \v{S}trobl,$^{11}$ Andrei V. Berdyugin,$^2$}
\newauthor{Stefano Ciprini,$^{12,13}$ Daniel E. Reichart,$^{14}$ Vladimir V. Kouprianov,$^{14}$}
\newauthor{Katsura Matsumoto,$^{15}$ Marek Drozdz,$^{16}$ Markus Mugrauer,$^{17}$ Alberto Sadun,$^{18}$}
\newauthor{Michal Zejmo,$^{19}$ Aimo Sillanp\"a\"a,$^2$ Harry J. Lehto,$^2$ Kari Nilsson,$^1$}
\newauthor{Ryo Imazawa,$^{21}$, Makoto Uemura $^{22}$ and James W. Davidson Jr. $^{23}$}
\\
$^1$ FINCA, University of Turku, FI-20014 Turku, Finland\\
$^2$ Tuorla Observatory, Department of Physics and Astronomy, University of Turku, FI-20014 Turku, Finland\\
$^3$ Astronomical Observatory, Jagiellonian University, ul. Orla 171, 30-244 Krakow, Poland\\
$^4$ Department of Astronomy and Astrophysics, Tata Institute of Fundamental Research, 4000005 Mumbai, India\\
$^5$ Aalto University, Mets\"ahovi Radio Observatory, Mets\"ahovintie 114, 02540 Kylm\"al\"a, Finland\\
$^6$ Aalto University, Department of Electronics and Nanoengineering, P.O. Box 15500, FI-00076 AALTO, Finland\\
$^7$ Aryabhatta Research Institute of Observational Sciences (ARIES), Manora Park, Nainital 263001, India\\
$^8$ Nordic Optical Telescope, Apartado 474, E-38700 Santa Cruz de La Palma, Spain\\
$^9$ Instituto de Astrofisica de Andalucia - CSIC, Glorieta de la Astronomia s/n, 18008 Granada, Spain\\
$^{10}$ Czech Technical University, Faculty of Electrical Engineering, 16000 Prague, Czech Republic\\
$^{11}$ Astronomical Institute (ASU CAS), 251 65 Ond\v{r}ejov, Czech Republic\\
$^{12}$ Instituto Nazionale di Fisica Nucleare (INFN) Sezione di Roma Tor Vergata, Via della Ricerca Scientifica 1,     \\00133, Roma, Italy\\
$^{13}$ ASI Space Science Data Center (SSDC), Via del Politecnico, 00133, Roma, Italy\\
$^{14}$ University of North Carolina at Chapel Hill, Chapel Hill, North Carolina, NC 27599, USA\\
$^{15}$ Astronomical Institute, Osaka Kyoiku University, 4-698 Asahigaoka, Kashiwara, Osaka, 582-8582, Japan\\
$^{16}$ Mt. Suhora Observatory, Pedagogical University, ul. Podchorazych 2, 30-084 Krakow, Poland\\
$^{17}$ Astrophysikalisches Institut und Universitäts-Sternwarte, Schillergässchen 2, D-07745 Jena, Germany\\
$^{18}$ Department of Physics, University of Colorado, Denver, CO 80217, USA\\
$^{19}$ Kepler Institute of Astronomy, University of Zielona Gora, Lubuska 2, 65-265 Zielona Gora, Poland\\
$^{20}$ Key Laboratory for Research in Galaxies and Cosmology, Shanghai Astronomical Observatory,\\ Chinese Academy of Sciences, Shanghai 200030, China\\
$^{21}$ Department of Physics, Graduate School of Advanced Science and Engineering, Hiroshima University,\\ 1-3-1 Kagamiyama, Higashi-Hiroshima, Hiroshima 739-8526, Japan\\
$^{22}$ Hiroshima Astrophysical Science Center, Hiroshima University, 1-3-1 Kagamiyama, Higashi-Hiroshima, Hiroshima 739-8526, Japan\\
$^{23}$ Department of Astronomy, University of Virginia, 530 McCormick Rd., Charlottesville, VA 22904, USA
}

\date{Accepted ... Received ... in original form ...}

\pubyear{2022}

\begin{document}
\label{firstpage}
\pagerange{\pageref{firstpage}--\pageref{lastpage}}
\maketitle



\begin{abstract}

The highly variable blazar OJ~287 is commonly discussed as an example of a binary black hole system. The 130 year long optical light curve is well explained by a model where the central body is a massive black hole of 18.35$\times$10$^9$ solar mass that supports a thin accretion disc. The secondary black hole of 0.15$\times$10$^9$ solar mass impacts the disc twice during its 12 year orbit, and causes observable flares. Recently, it has been argued that an accretion disc with a typical AGN accretion rate and above mentioned central body mass should be 
at least six magnitudes brighter than OJ~287's host galaxy and would therefore be observationally excluded. Based on the observations of OJ~287's radio jet, detailed in Marscher and Jorstad (2011), and up-to-date accretion disc models of Azadi et al. (2022), we show that the V-band magnitude of the accretion disc is unlikely to exceed the host galaxy brightness by more than one magnitude, and could well be fainter than the host. This is because accretion power is necessary to launch the jet as well as to create  electromagnetic radiation, distributed across many wavelengths, and not concentrated especially on the optical V-band. Further, we note that the claimed V-band concentration of accretion power leads to serious problems while interpreting observations of other Active Galactic Nuclei.
Therefore, we infer that the mass of the primary black hole and its accretion rate do not need to be smaller than what is determined in the standard model for OJ~287.  
\end{abstract}


\section{Introduction} \label{sec:intro}

OJ~287 has the longest and best covered optical light curve among all AGN (Active Galactic Nuclei). It has been photographed as a part of other projects since 1888, and for the last 50 years it is among the most studied extragalactic objects. The number of optical observations in the light curve now exceeds 100,000. The light curve is shown in Figure 1. 
A binary black hole (BH) model for OJ~287 was proposed in 1988 \citep{sil88}. The model and its subsequent elaborations \citep{LV96,val07,Dey18} have made increasingly accurate predictions about future light curve events, in 1994, 1995, 2005, 2007, 2015, and 2019 which have been subsequently verified \citep{sil96a,sil96b,val06,val08,val16,Laine20,val21}. The multi-wavelength  observations at these times as well as the study of the historical light curve as a whole, has produced an increasingly accurate picture of the central engine of OJ~287 (Figure 1). 
At the heart of this model is a massive BH that weighs more than $10^{10}$ solar mass. It is customary to refer such massive black holes 
as ultramassive black holes, even though they simply represent a continuum distribution which extends well beyond the mass of the BH in OJ~287 \citep{sag16}.  

Recently, \cite{Komossa2023} (K23 hereafter) claimed that the brightness of the accretion disc in OJ~287 should be at least V $\sim$ 13.5 in the V-band. Looking at Figure 1, we obviously do not see such a constant contribution from the accretion disc. Therefore K23 conclude that the standard model has a mass which is at least two orders of magnitude too high, or its accretion rate is similarly two orders of magnitude too high. In the latter case the disc should not be geometrically thin which in their opinion invalidates the standard model.

\begin{figure}
\centering
\includegraphics[scale=0.074]{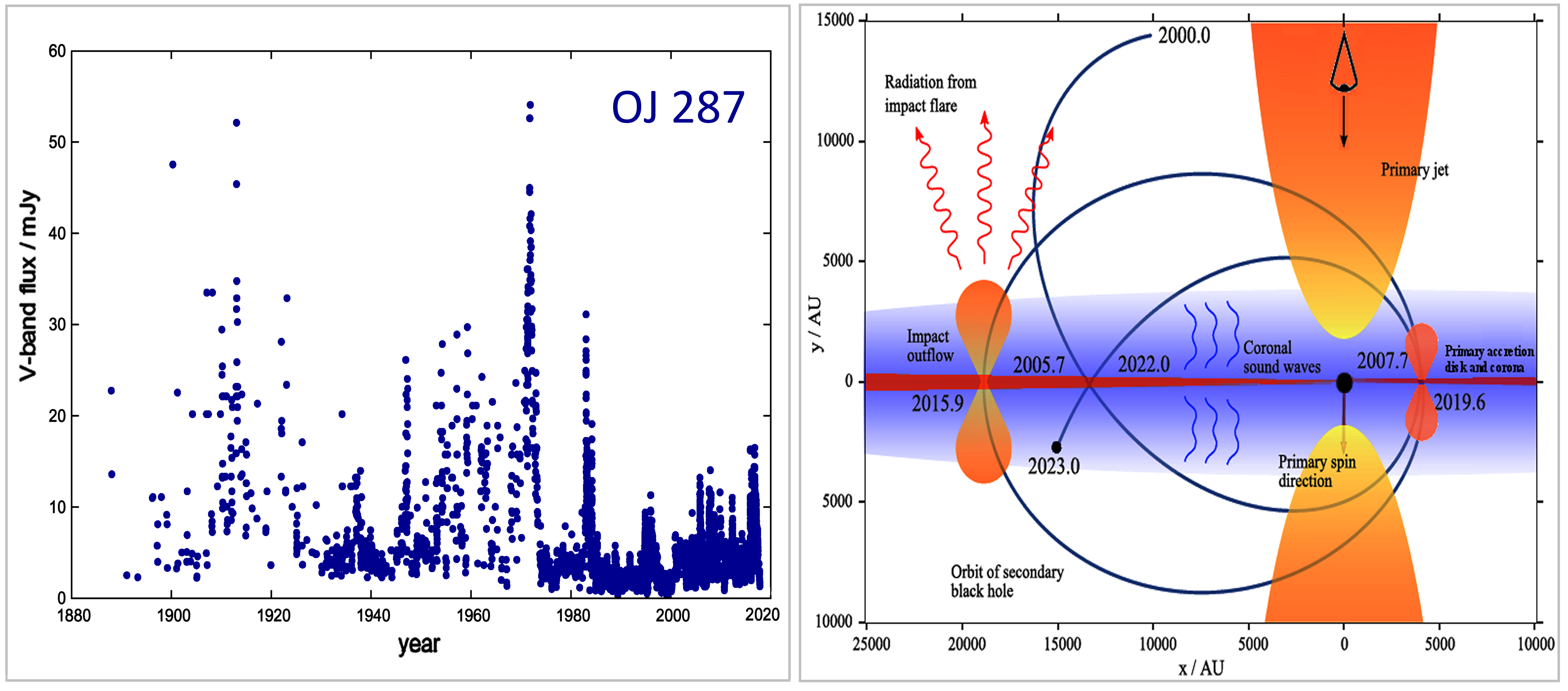}
\caption{The historical light curve of OJ~287 in the optical V-band (on the left) and the binary black hole picture of its central engine (on the right). The model has a 18.35$\times$10$^9$ solar mass central BH, with an accretion disc with the accretion rate of 8$\pm$4\% of the Eddington rate. According to \citet{Komossa2023}, an accretion disc of these parameters should be as bright as 15 mJy in the V-band,  which is clearly not seen, as it would provide the bottom level of flux in this graph. In the standard accretion disc models the disc brightness is orders of magnitude lower, and therefore the disc cannot be observed in the light curve.}
\label{fig1}
\end{figure}

 K23 base their arguments on calculating the total accretion power of OJ~287 and channelling a major part of it to the V-band. They assume (1) that 100 percent of the accretion power appears as accretion disc radiation, (2) that 1/9 of the disc radiation goes to the optical wavebands (so called bolometric correction, \citep{kas00}), and (3) that 100 percent of the optical radiation is concentrated in the V-band. In this way they calculate the total emission $L_V\sim 2 \times 10^{46}$ ergs sec$^{-1}$. Then they convert this emission rate to the V band magnitude corresponding to the distance of OJ~287 \citep{nil10}, and find that it is at least 10 times higher than the recent minimum brightness observed in 2022. This means that the disc should at least 6 magnitudes brighter than the host galaxy \citep{nil20}.
 
 It is straightforward to note that the above estimate provides
 an  upper limit for OJ~287's disc radiation, 
 and thus gives a lower limit for V band magnitude.
 We note that  K23's assumption (2) is consistent 
 with the detailed studies on various accretion disc models by 
\cite{aza22} though their assumptions (1) and (3) can lead to a huge overestimate for OJ~287's accretion disc energy flux.
 
 In this paper, we provide realistic estimates for the  upper limit for
 OJ~287's disc luminosity, influenced by the fact that there 
 is no observational evidence for the accretion disc in OJ~287.
  
 For this purpose, we employ recent detailed models of accretion discs by \cite{aza22} and take into account the observed radio jet in our energy budget estimates \citep{mar11}, while considering the fact that power can also be dissipated into accretion disc winds \citep{rod20}. 
 
 Thereafter, we explore consequences of the K23-type  
 estimates while dealing with AGNs in the nearby Universe 
 where we have much more detailed observational inputs compared
 to OJ~287.
 
\section{ Estimating Accretion Disc Luminosities in AGNs}
\label{sec:consider}

The calculation of the accretion disc structure and its bolometric corrections is not a simple task and has required lots of work in modelling as well as in comparing with observations \citep{sha73,min90,che95,zdz98,dek99,col01,nem10,aza22}. The recent models of \cite{kub18} and \cite{aza22} give us the luminosity of the disc as a function of the main parameters of the system, the mass of the central BH $M_{BH}$, the accretion rate $\lambda_{Edd}$ with respect to the the rate that produces the Eddington luminosity, and the spin of the black hole. For the OJ~287 primary we have found the central mass $M_{BH}=18.35\times10^9$ solar mass, accretion rate $0.08\pm0.04$ of the Eddington limit and the normalised spin $\sim$ 0.38 \citep{Dey18,val19}.

The Eddington ratio is defined as the mass accretion rate, $\dot{m}$, normalised by the rate that produces the Eddington luminosity, $\dot{m}_{Edd}$, as:
\begin{equation}
\lambda_{Edd} = \dot{m} / \dot{m}_{Edd},
\end{equation}
where
\begin{equation}
\dot{m}_{Edd}  = L_{Edd} / (\eta_r  c^2),
\end{equation}
and $\eta_r$ is the radiation efficiency, usually taken as $\eta_r = 0.057$. The models of \cite{kub18} and \cite{aza22} as well as our standard model \cite{Dey18} use this definition \citep{ste84,val19,jia19}.

In magnetic disc models a large fraction of the accretion power goes to maintaining the hot corona (about $50\%$ in the model AGN0.07 of \cite{jia19}, representing $\lambda_{Edd,corona} \sim 0.035$), mechanical power of the disc wind (about $40\%$ in the model of \cite{rod20}, representing $\lambda_{Edd,wind} \sim 0.048$), and the mechanical power of the twin jets ($\sim 5\times10^{46}$ erg sec$^{-1}$ from the observed megaparsec jet \citep{mar11}, representing $\lambda_{Edd,jet} \sim 0.033$ for the twin jets). Added together, these magnetic and mechanical contributions could in principle soak nearly all the accretion energy in our model, where $\lambda_{Edd,model} = 0.08 \pm 0.04$.

In the \cite{aza22} grid of models this puts us in the area slightly beyond $M_{BH}=10^{10}$ solar mass, log$(\lambda_{Edd})$ smaller than -1.5, and the spin closer to 0 than 1. Figure 5 (top left panel) of \cite{aza22} gives log$(\nu L_{\nu}) \sim 45.8$ where $\nu L_{\nu}$ is in units of ergs sec$^{-1}$. Extrapolation from Figure 7 (top right panel) of \cite{aza22} to log$(\lambda_{Edd}) \sim -2.5$ gives $\nu L_{\nu} \sim (0.5 - 2)\times10^{45}$ ergs sec$^{-1}$. Considering the large fraction of the accretion power likely to go to the halo, winds and jets, this may be a realistic estimate for the power that may appear as disc luminosity.

Out of the disc luminosity (1 - 4)$\times$10$^{44}$ ergs sec$^{-1}$ is shared into the V-band. This corresponds roughly to V$\sim 19\pm1$. It is close to the magnitude of the host galaxy measured directly \citep{nil20}, and agrees with the V$\sim$17.5 observational upper bound at the historical brightness minimum when the disc was definitely not detected, based on the colours, unless the disc mimics the colours of the host galaxy \citep{tak90,val22}. 
 
OJ~287 is not the optimal system for testing accretion disc theories since in this system the disc is overpowered by the jet. Therefore we may ask where does the K23-type calculation lead when applied to other AGN. 

We first use empirical relations to estimate the AGN contribution in the V-band lumininosity of OJ~287. \cite{jal23} find that for an AGN around redshift 0.3 and Eddington ratio around 8 $\%$, the AGN and host galaxy luminosities are approximately equal, with a wide variation in individual cases. This agrees with what we just found out for OJ~287.

If a central BH of $M_{BH}\sim 1.835 \times 10^{10}$ solar mass and a typical accretion rate makes the accretion disc 6 magnitudes brighter than the host galaxy, what would be the implications to other accretions discs with lower BH mass? The BH mass has a nearly linear correlation with the host luminosity \citep{sag16}, and since the accretion disc power also correlates directly with the BH-mass \citep{aza22}, in close to linear manner, we have to conclude that also in general the accretion discs should be brighter than their host galaxies. 

In Figure 2 we plot the BH mass - host galaxy mass diagram for the local AGN. It is the usual diagram \citep{sag16,ter17} for well determined BH masses, but including only those galaxies which have been reported to possess an AGN. Then we draw a line which runs two orders of magnitude below the observed point for OJ287 \citep{nil20}, and has a slope from the above mentioned correlations. Therefore in the K23-calculation it represents the line of equality between the accretion disc and the host galaxy. The observational points are not far from this line which means that all AGN in the nearby Universe should be about equally bright as their hosts, and many of them much brighter than their hosts.

\begin{figure*}
\includegraphics[scale=0.81, angle=0]{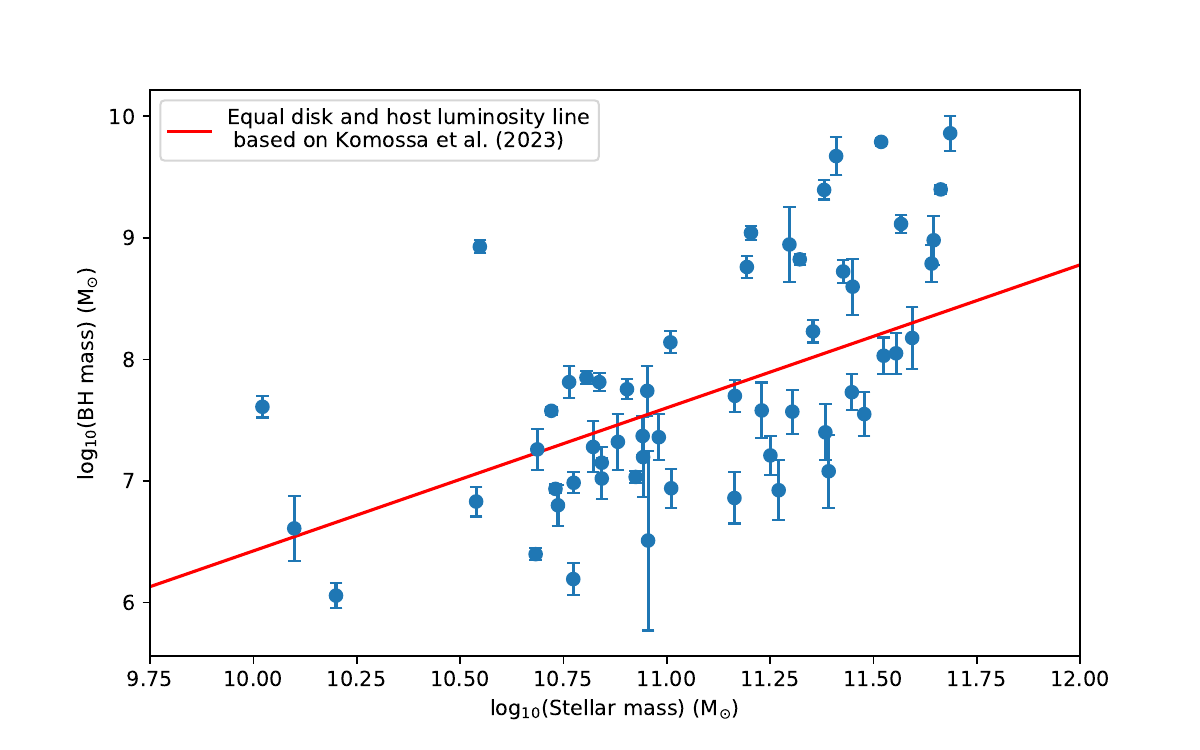}
\centering
\caption{The correlation between black hole mass and the host mass in nearby AGN (crosses) \citep{sag16,ter17}. We start from the K23 estimate that the brightness of a 1.835$\times10^{10}$ solar mass BH accretion disc is V = 13.5 at the redshift 0.306 of OJ~287, given the accretion rate of 8$\%$ of the Eddington rate. Then we use the disc luminosity - BH mass correlation from \citet{aza22} to transfer the estimate to the AGN of the local Universe, and draw a line representing systems where the accretion disc luminosity should be equal to the host galaxy luminosity in the K23 accretion disc calculation. We see that the observational points concentrate around the line, suggesting that the accretions discs should be similar in brightness to the host galaxies. Also a large number of local AGN should overpower their host galaxies by more than an order of magnitude. However, observationally we know that this is not the case, and thus the K23-calculations are oversimplified.}
\label{oj287fig2}
\end{figure*}

From observations we know that this is not the case: the nearby AGN do not dominate their host galaxies. Typically it is difficult to see the accretion disc, and even if we see it, it is definitely very much fainter than host galaxy \citep{sun89}. A way to get around this problem in the K23 model is to reduce the accretion rate from $\sim8\%$ to a very much lower value, for all AGN, but that does not seem to be possible \citep{kyn23}.

In quasars it has been found that accretion discs may be up to ten times brighter than their host galaxies in the optical wavebands \citep{sun89}. The discs are thought to be responsible for the "Big Blue Bump", which dominates the optical luminosity in these objects \citep{col01}. Even here the K23-type calculation gives a one magnitude excess of the disc luminosity, and considering that quasars often accrete close to their Eddington rates, much higher than the $8\%$ in the K23 calculation, the problem becomes even worse. It is possible that also all quasars contradict the K23-type calculation.

\section{Discussion and conclusions}

One may wonder what is the origin of the difference of 5.5 magnitudes of the V-band luminosity of OJ~287 between the K23 calculation and our calculation based on recent disc models. About one half of the difference (in magnitude scale) appears to come from the fact that the K23 model concentrates the accretion power very strongly on the V-band. The other half has to do with the jet: the mechanical power required by the jet is very large \citep{mar11}, and this energy cannot appear as disc luminosity.  In addition, the mechanical luminosity of the AGN wind in OJ~287 may carry away up to 50 $\%$ of the mass accretion power \citep{rod20}. Therefore a calculation based on the theoretical accretion power, and assuming that the disc has to radiate all of this, provides an upper limit on the disc luminosity at best, and as we have seen, this upper limit is not very useful.

Also we have to remember that the accretion rate in the standard model of \cite{Dey18} is not solved directly, but it comes via a further calculation from the parameters of the orbit solution \citep{val19}. Therefore it has wide error bounds which have to be taken into consideration if one attempts to estimate the upper bound on the disc luminosity.

Our estimate for the accretion disc brightness in OJ~287 is V$=19\pm1$. The upper limit of brightness, corresponding to V=18, is consistent not having detected the disc during the faintest state ever seen in OJ~287 \citep{tak90}. Or possibly, if the accretion disc has very similar colours as the host host galaxy, it could have been detected at the level of V=18 \citep{val22}. 

To resolve this problem, K23 proposed to lower the central BH mass to $\sim 10^8$ solar mass. However, if there is really a problem with the disc brightness, changing the BH mass is not a solution. The power of the jet must still be supported, and it depends on the same product of mass and mass accretion rate as the disc brightness. If mass is changed, the accretion rate must be adjusted equally, but in opposite direction, to keep the product constant. The product is fixed by the jet power \citep{mar11}, and so is the disc brightness also.

The suggestion of lowering the BH mass to $\sim 10^8$ solar mass rather brings problems of its own. It would imply that the accretion rate of OJ~287 would have to be highly super-Eddington even to produce the mechanical power of the jet, and from the great length of its jet we may deduce that it would have been in a super-Eddington state for a long time \citep{mar11}. This is rather unlikely \citep{jal23}. This is a problem of the low BH mass scenario both in binary and single BH models of OJ~287 \citep{bri18}.

Further problems arise, if OJ~287 is interpreted as a low BH mass binary system. The usual interpretation is based on the double peaked structure of the optical light curve. The simplest way to obtain the double peaks in a binary system is to associate the observed flares with disc impacts twice during each orbit. But lowering the mass scale by two orders of magnitude lowers also the brightness of the flares by the same amount, and makes them invisible against the background of the general jet emission. This applies equally to direct impact flares and flares generated by tidally induced accretion flow variations \citep{val12}. This of course does not exclude the possibility that a new way of producing double peaked structure is invented in future.

The main unanswered question in the low BH mass scenario is why OJ~287 behaves in a highly predictable manner in the standard model of \citet{Dey18}. The standard model relies heavily on General Relativity, and the latter is not a scale-free theory in the same sense as the Newtonian theory is. Therefore the predictability, which has now been witnessed for 40 years \citep{val23}, is lost if the mass scale is lowered.

There were two major flares predicted to take place in OJ~287 also in 2022, the first one in January/February and the second one in July/August \citep{val21}. Because OJ~287 is close to the sun during the latter period, ground based optical observations are not possible, and thus the 2022 observing campaign carried out by us was concentrated on the first half of the year. The first flare is caused directly by the impact of the secondary BH on the disk. A cloud of plasma is pulled out of the  disc \citep{iva98}, and its radiation may be described by the standard \cite{van71} expanding cloud model. The model provides an explanation for the peculiar behaviour of the flare at the time when the source becomes optically thin to synchrotron radiation.

The flare was observed at the predicted time, with properties that have been observed once before in OJ~287, in 2005 \citep{ciprini2008}. In 2005 and in 2022 the impacts took place at practically the same distance from the primary BH which should make the impact flares comparable to each other. The time difference between the 2005 and 2022 flares was expected to be 6161$\pm$3 days from the \cite{Dey18} orbit model, while the observed distance between the flat spectrum peaks was within the same time interval. In 2005, we know from the orbit model that the flare peaked 35$\pm3$ days after the disc impact. In 2022 we also have strong evidence for the association with the disc impact, with the same time delay from the impact to the flat spectrum peak. The light curves of both flares are displayed in \cite{val23}. The configuration of the emitting cloud at the brightness peak is illustrated by \cite{iva98} (Figure 4, top right hand panel).

As to the July/August flare, it arises when the plasma cloud has expanded further by about a factor of ten, and the cloud becomes optically thin to bremsstrahlung radiation. At the start of our observing campaign, the impact configuration had not been calculated yet, and therefore there was a real possibility that the flare could have shifted outside the "summer gap", and this possibility was communicated also to wider circles (a preprint posted in arXiv).



However, once the configuration was determined, it became clear that the second flare was unobservable, and indeed, was not observed. Since we could not observe the second flare, we can only say that its absence in our light curve is consistent with our model \citep{val07,Dey18,val21,val23}.  

\section*{Acknowledgements}
This work was partly funded by NCN grant No. 2018/29/B/ST9/01793 (SZ) and JSPS KAKENHI grant No. 19K03930 (KM). SC acknowledges support by ASI through contract ASI-INFN 2021-43-HH.0 for SSDC, and Instituto Nazionale di Fisica Nucleare (INFN). RH acknowledges the EU project H2020 AHEAD2020, grant agreement 871158, and internal CTU grant SGS21/120/OHK3/2T/13. ACG is partially supported by Chinese Academy of Sciences (CAS) President's International Fellowship Initiative (PIFI) (grant no. 2016VMB073). MJV acknowledges a grant from the Finnish Society for Sciences and Letters.

\section*{Data Availability}
We do not present previously unpublished data in this work.

\bsp    
\label{lastpage}
\end{document}